\documentclass[conference]{IEEEtran}
\IEEEoverridecommandlockouts
\usepackage{cite}
\usepackage[T1]{fontenc}
\usepackage{amsmath,amssymb,amsfonts}
\usepackage{algorithmic}
\usepackage{graphicx}
\usepackage{textcomp}
\usepackage{xcolor}
\usepackage[hyphens]{url}
\usepackage{pdfpages}
\def\BibTeX{{\rm B\kern-.05em{\sc i\kern-.025em b}\kern-.08em
    T\kern-.1667em\lower.7ex\hbox{E}\kern-.125emX}}
\begin{document}

\title{Full-text Search for Verifiable Credential Metadata on Distributed Ledgers\\
\thanks{This work has been conducted within the framework of the Decentralized Identity Management System (DIMS) project in the focus area of Digital Finance, funded by EIT Digital.}
}

\author{
\IEEEauthorblockN{Zolt\'{a}n Andr\'{a}s Lux, Felix Beierle, Sebastian Zickau, and Sebastian G\"{o}nd\"{o}r}
\IEEEauthorblockA{\textit{Service-centric Networking} \\
\textit{Telekom Innovation Laboratories}\\
\textit{Technische Universi{\"a}t Berlin}\\
Berlin, Germany \\
\{ zoltan.a.lux | beierle | sebastian.zickau | sebastian.goendoer \}@tu-berlin.de}
}

\maketitle

\begin{abstract}
Self-sovereign Identity (SSI) powered by distributed ledger technologies enables more flexible and faster digital identification workflows, while at the same time limiting the control and influence of central authorities. However, a global identity solution must be able to handle myriad credential types from millions of issuing organizations. As metadata about types of digital credentials is readable by everyone on the public permissioned ledger with Hyperledger Indy, anyone could find relevant and trusted credential types for their use cases by looking at the records on the blockchain. To this date, no efficient full-text search mechanism exists that would allow users to search for credential types in a simple and efficient fashion tightly integrated into their applications. In this work, we propose a full-text search framework based on the publicly available metadata on the Hyperledger Indy ledger for retrieving matching credential types. 
The proposed solution is able to find credential types based on textual input from the user by using a full-text search engine and maintaining a local copy of the ledger.
Thus, we do not need to rely on information about credentials coming from a very large candidate pool of third parties we would need to trust, such as the website of a company displaying its own identifier and a list of issued credentials. We have also proven the feasiblity of the concept by implementing and evaluating a prototype of the full-text credential metadata search service.
\end{abstract}


\section{Introduction}
\label{Introduction}

A prominent application area of blockchain technologies is digital identity, where numerous organizations are interested in establishing a decentralized ecosystem\footnote{Decentralized Identity Foundation Website: https://identity.foundation, accessed on 28.8.2019}. Distributed ledger based decentralized identity solutions give more control in the hands of the identity holder in what information they want to disclose to other parties. The Sovrin whitepaper \cite{Whitepaper:Sovrin} cites that \textit{"one seventh of the world's population has no legal identity"}, blockchain powered identity systems have the potential of becoming the foundation of a globally available, privacy aware, accessible by everyone means of identification while still maintaining trust among the participants. This would be a major game changer for identity holders, as the identity landscape nowadays is centralized and controlled by a few large players like Facebook and Google. End users do not have control over their own identity, and customer data is sold with little transparency and concern for the privacy of the end user. Central identity providers are also attractive targets for attacks, as hackers have the chance of getting sensitive data about potentially hundreds of millions of individuals by being able to infiltrate a single system.

When thinking about a user-centric identity framework we would emphasize security, usability, privacy, and consent at the same time. With self-sovereign Identity (SSI) the end users (identity holders) have the control over their identity, that no single central entity can take away. 

The SSI movement aims to introduce privacy, security, accessibility, and transparency by decentralization into today's digital identity landscape. The 10 principles of SSI formulated by Christopher Allen include consent, protection, control, transparency, portability, interoperability, and minimalization \cite{allen2016path}. The Sovrin whitepaper by Tobin and Reed \cite{tobin2016inevitable} followed and referenced Allen's original blog entry \cite{allen2016path}, which envisioned the Sovrin Identity Network.

\begin{figure*}[t!]
    \centering
    \includegraphics[trim={1.2cm 22.25cm 3.25cm 1.1cm},clip, width=\textwidth]{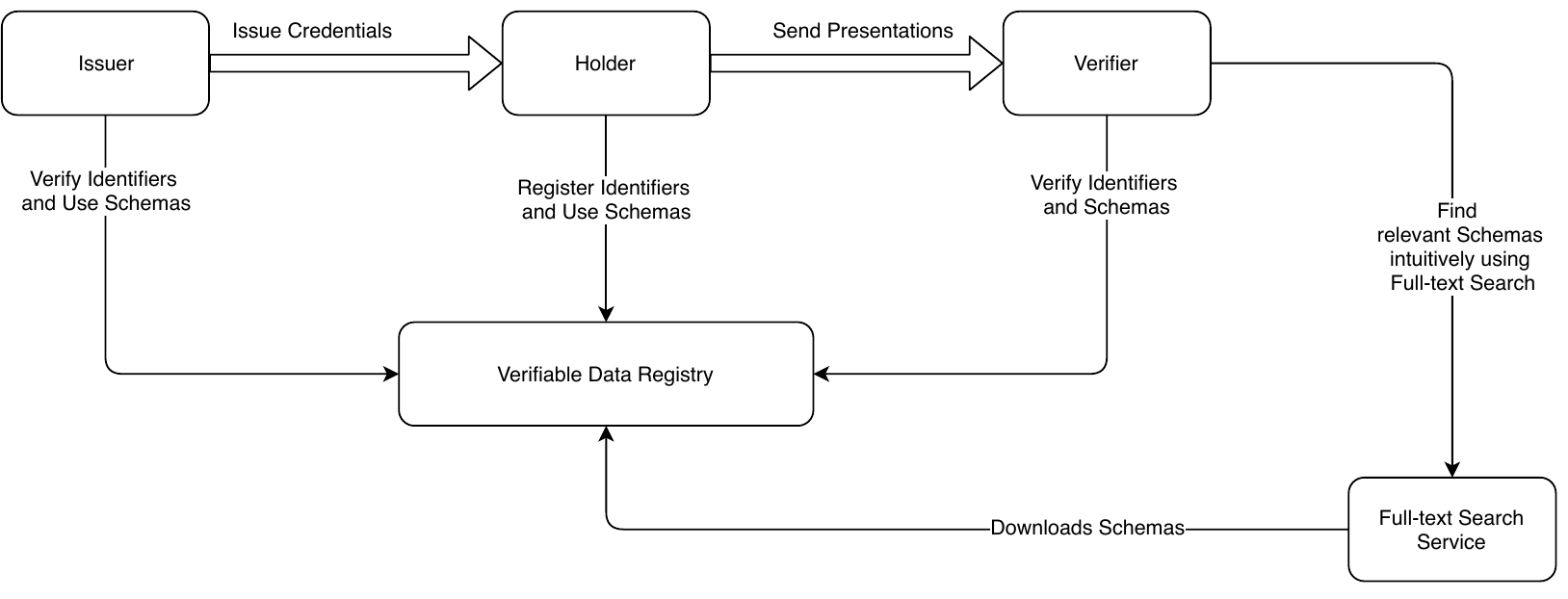}
    \caption{Verifiable Credentials Role Model and Search System Integration. The three main roles are: (1) issuer , (2) holder and (3) verifier. Another essential component of the model is a verifiable data registry. The issuer sends credentials to the holder. Then, the holder can send presentations of the received credentials to a verifier. The search service relies on the publicly readable Verifiable Data Registry. Source of the Verifiable Credentials Role model: W3C \cite{Webpage:VC}}
    \label{fig:ssi_search}
\end{figure*}

The SSI role model comprises an issuer, an identity holder, and a verifier, as depicted in Figure \ref{fig:ssi_search} along with our solution for verifiable credential metadata search on the ledger. The heart of an SSI solutions can be a public permissioned distributed ledger, which serves as a decentralized, tamper-proof "verifiable data registry" \cite{Webpage:VC}, readable by everyone. For the sake of privacy and scalability, only minimal information is stored on the ledger, thus minimizing the publicly available data to essential information that needs to be accessible for all participants in the network. The empty templates of verifiable credentials and public keys for verification are stored on the distributed ledger with Hyperledger Indy, because in the SSI vision everyone needs to be able to verify the validity of a received credential. For example, we can think about storing the attribute names of an ID card like name, date of birth, etc. and the respective public key along with the name of the credential (e.g. ID card) and the identifier of the government. Another illustration of identity with blockchain and distributed ledgers would be to think of information about credentials types on the ledger as the data definition language for a table in the context of relational databases and SQL, where the actual records of the table are each stored by the identity holders.  

Applying the SSI workflow is in many use cases non-trivial, as we can think about multiple valid identification scenarios, like verification at the website of a bank or a government office. 

Furthermore, there are also numerous ways to identify ourselves with various types of credentials, and verifiers need to be able to formulate restrictions about their preferred types of credentials and/or issuers for saving time by not having to check received documents individually and having a higher degree of automatization.

When developing applications powered by SSI, we have faced the verifier's problem of spotting the right credentials among the information stored on the ledger, thus they can request for proof for attributes from relevant credentials. It is necessary for the verifier side to be able to find trustworthy information about credentials relevant for their use cases, thus they can decide which credentials to accept without having to check each credential definition individually.

We propose an intuitive full-text ledger search service for solving this problem, which enables administrators, developers and possibly users of applications to leverage the benefits of verifiable credentials by finding necessary information on the ledger similar to how we search for relevant pages on the world wide web.

For example, a company offering housing to students on its website would like to recieve proofs of enrollment at universities nearby, thus saving precious time by reducing manual checking. This kind of restriction can be easily configured and maintained with our full-text ledger search by looking up relevant types of credentials found on the Indy ledger with textual queries.

Furthermore, issuers could also leverage the full-text search service for finding Hyperledger Indy schemas from other entities they want to base their credential definitions on. 

Thus, the main contribution of our paper is (a) that we have identified the need for a tightly integrated full-text ledger search feature for flexible and usable applications utilizing verifiable credentials, and (b) implemented it using Apache Solr \cite{Webpage:ApacheSolr} for Hyperledger Indy and evaluated its performance in order to achieve a flexible and efficient SSI workflow in the current dynamically evolving landscape of blockchain technologies.

The rest of the paper is structured as follows: Section \ref{related_work} provides an extensive overview of the state of the art in SSI, Section \ref{design} and Section \ref{implementation} we describe design and implentation specific details of our full-text search system, Section \ref{evaluation} evaluates the verifiable credential metadata search service from various perspectives, and Section \ref{conclusion} is about conclusion and future work.

\section{Related Work}
\label{related_work}

Bitcoin \cite{nakamoto2008bitcoin}, the first blockchain, was invented for creating a peer-to-peer electronic cash system. The technology soon became a highlight in the tech realm, and it inspired numerous decentralized blockchains such as Ethereum \cite{wood2014ethereum}. Nowadays blockchain and distributed technologies are used for other use cases in addition to electronic payment, such as digital identity and supply chains.

As Dunphy and Petitcolas \cite{dunphy2018first} write, Namecoin\footnote{Namecoin Website: https://www.namecoin.org/, accessed on 28.8.2019}, a cryptocurrency based name server for the .bit domain, pioneered in creating "a secure and decentralised identifier-attribute mapping" using distributed ledger technologies, which contradicts what is known as Zooko's Traingle \cite{wilcox2003names}, that a naming system can only satisfy 2 of the following 3 criteria: (1) secure, (2) decentralized and (3) human-readable.

Certcoin \cite{fromknecht2014decentralized} already aims at establishing a fully-fledged Public Key Infrastructure with retention using Blockchain, meaning that impersonation of an already registered user is impossible.

The Decentralized Public Key Infrastructure (DPKI) is depicted in a whitepaper \cite{allen2015decentralized} co-authored by Vitalik Buterin, Christopher Allen, Markus Sabadello, Drummond Reed and others, that describe the current model of digital identity controlled by third parties \textit{"results in severe usability and security challanges Internet-wide"} \cite{allen2015decentralized}, and they argue that DPKI addresses many of these.

The creation of anonymous credentials with blockchain is also possible \cite{garman2014decentralized}.

Today, there are already numerous organizations \cite{Webpage:DIF} who are looking forward to leverage the potential advantages of distributed ledger technologies for the digital identity use cases \cite{Webpage:VCUseCases} at a global scale. Still, identity is also supposed to be interoperable, and Decentralized Identifiers (DIDs) and verifiable credentials are being standardized by the world wide web consortium.

The Bundesdruckerei in Germany is already running a Demo-B\"{u}rgeramt pilot project \cite{Webpage:BUDRU}, where we can receive digital ID card, driving license, boatmasters' certificate, student ID, event ticket, or parking permit verifiable credentials. The website lets us observe an example of a simple identification workflow using our Jolocom wallet app.

A survey from M{\"u}hle \cite{muhle2018survey} provides a good overview of the motivation behind SSI. As the SSI terminology has evolved during the last years, we would recommend the variants from the W3C working group \cite{Webpage:VC}, \cite{Webpage:DID} as a point of reference.

\subsection{Identification Workflow}

The workflow of identification with verifiable credentials fits a wide range of scenarios. For example, we can think of a completely online process, such as opening an account on the website of a bank, or proving our age for ordering a drink in a bar using a zero-knowledge proof generated by an SSI wallet app on our smartphone. In an interactive and flexible setting, the verifier should be able to retrieve information about different ID card issuers for creating a proof request suitable for the specific scenario. By designing a flexible yet trusted identification workflow we can address more use cases, thus enabling wider usage of SSI solutions.

The preferred use case for this paper is that the verifier's side is using the full-text search, however it could potentially be just as necessary for an issuer creating credentials. We can think of multiple scenarios where the verifier's side could benefit from using the full-text search. 

Primarily, an administrator or developer of a verifier application can use the full-text search for specifying restrictions about accepted credentials in an intuitive and user friendly way, so hard-coding restrictions about credentials would not be needed, which improves maintainability and reduces development costs and  saves time. One of several appealing specific application scenarios for the full-text verifiable credentials search is an online housing platform enhanced with trusted data through SSI, where a large number of restrictions about acceptable verifiable credentials needed to be configured like proofs of not having rent arrears, proof of income, credit score and an identity document for being able to apply for a flat. There are often more specific requirements like having a student status for being eligible for an apartment in a dormitory close to the university. A student could also get a discount upon showing a digital proof of matriculation when registering online for membership at a local swimming club.

Thus, we identify a full-text ledger search service as a key feature when developing applications leveraging the trust benefits of SSI for achieving flexible, maintainable and efficient business processes utilizing decentralized identifiers and verifiable credentials.

\subsection{Decentralized Identifiers}

Self-sovereign identity introduces DIDs, controlled solely by the identity holder. DIDs enable more privacy for the identity holder, as anyone can have as many DIDs as they wish, thus, someone can use a distinct DID for each connection, mitigating the risk of being correlated. DIDs are currently being standardized by the World Wide Web Consortium \cite{Webpage:DID}. Several organizations are developing their own DID methods \cite{Webpage:DIDMethods}, and many of these organizations are startups, such as Jolocom \cite{Webpage:Jolocom} or Ocean Protocol \cite{Webpage:OceanProtocol}. The SSI community also aims to achieve interoperability, and there is already an open source source software, called Uniresolver \cite{Webpage:Uniresolver} with the source code available on  GitHub \cite{GitHub:Uniresolver}, which is able to resolve DIDs to the respective DID Document, and anyone is free to develop drivers for their own DID methods. There is also a software for registering DIDs, which is called Uniregistrar \cite{Webpage:Uniregistrar}.
DIDs are not only meant as identifers for humans, but also for IoT devices or even data \cite{Webpage:DIDEverything}.

Other forms of decentralized identifiers have been proposed in the past. In \cite{gondor2016distributed} and \cite{gondor2018seamless}, a distributed globally unique identification framework is proposed that uses hashes of public keys to identify users. The solution implements a DHT-based directory service that manages discovery of user profiles and identity management and therefor allows users to independently create and register identities.
To benefit from advantages of DLT, similar, even more flexible solutions have been proposed. The Ethereum Name Service (ENS\footnote{ENS Website; \url{https://ens.domains/}, accessed on 21.7.2019}) creates a link between human readable identifiers and attributes such as address or name. Similar to the Domain Name System (DNS), top-level domains in the ENS are owned by registrar contracts and may have sub-domains. Each domain entry in a registrar references a separate resolver contract that stores the associated information, such as an IP address. While the ENS proposes a default resolver, specialized custom resolver contracts may define additional attributes. Similar concepts have been implemented by systems such as Namecoin \cite{Webpage:Namecoin} or Blockstack \cite{ali2016blockstack}.

\subsection{Verifiable Credentials}

Verifiable credentials are essentially digital versions of physical credentials, but they are more tamper-evident by utilizing cryptography. Verifiable credentials are currently being standardized by the World Wide Web Consortium \cite{Webpage:VC}, and the specifications are already at an advanced stage.

\subsubsection{Sovrin and Hyperledger Indy} Sovrin \cite{Webpage:Sovrin} (envisioned by Dmitry Khovratovich and Jason Law \cite{khovratovich2017sovrin}) is an open source, distributed ledger based identity paltform specifically developed for SSI with a strong focus on the privacy of the identity holder. The Sovrin foundation contributed the initial codebase to the Hyperledger Indy project \cite{GitHub:Indy}, furthermore it is also an active member of the Hyperledger Aries \cite{Webpage:Aries} and Ursa \cite{Webpage:Ursa} projects. The Sovrin ledger, a deployment of Hyperledger Indy, is run and maintained by the stewards of the Sovrin foundation, who are responsible for maintaining trust in the Sovrin ecosystem. Stewards of Sovrin include Telekom Innovation Laboratories, IBM, Cisco, and several other organizations. Stewards can grant the role trust anchor to entities, and these entities have the right to write to the ledger. Writing to the ledger is necessary for being able to create verifiable credentials with Sovrin. Hyperledger Indy is a public permissioned ledger not linked to any cryptocurrency. However, there are plans to to introduce the Sovrin token. The stewards run a byzantine fault tolerant consensus algorithm \cite{lamport1982byzantine} called the plenum protocol \cite{Webpage:Plenum}. The Plenum protocol is based on the PBFT algorithm \cite{castro1999practical}. With dozens of nodes running the consensus algorithm (in case of Sovrin the Stewards), BFT algorithms can achieve a throughput of thousands of transactions per second \cite{buchman2016tendermint}.

An issuer needs to be a trust anchor, as they have to write the respective schema and credential definitions to the ledger. A schema has a name, a version and a number of attributes of string type without any additional restriction. A credential definition is created based on a schema, and it contains all the necessary public keys per attribute needed for verification. Issuers are able to create credential offers to identity holders based on credential definitions written on the ledger by them. The identity holder needs to accept the credential offer and send a credential request. Once the issuer has received the credential request, he/she is able to send the credential to the identity holder. Then the identity holder stores the credential in his/her wallet, and is able to generate proofs for each attribute of the credential. For the part of proving, the verifier and identity holder first need to establish a pairwise private connection. Then the verifier sends a proof request for a set of attributes from a number of credential definitions or schemas to a specific identity holder. Alternatively, the verifier can also specify the requirements about the issuer of the credential. Then the identity holder can generate a proof for attributes which are included in her credentials. The verifier can verify the received proofs of validity and non-revocation based on the credential definitions found on the ledger. Note, that Hyperledger Indy supports revocation implemented by the help of cryptographic accumulators to preserve the privacy of the user. It is worth mentioning, that an identity holder is encouraged to use a different DID for each connection, which makes correlation more difficult, and that these private DIDs of entities are not written on the public ledger by default.

The government of British Columbia created the Verifiable Organizations Network (VONX) \cite{Webpage:VONX} based on Hyperledger Indy. The Orgbook BC \cite{Webpage:Orgbook} lists 1.9 million verifiable organizations and 1.2 million active legal entities as of August 28, 2019.

According to Nadia Hewett (Project Lead for Blockchain, World Economic Forum, Centre for the Fourth Industrial Revolution), \textit{"In a data-driven, hyperconnected and digital world, businesses need a comprehensive system for trustworthy verification of other entities digital identity. British Columbia's Verifiable Organizations Network will empower businesses with a trusted digital identity issued by their local government. This is a prerequisite for digital optimization in the Fourth Industrial Revolution"} \cite{Webpage:Orgbook}.

\subsubsection{VONX Sovrin Mainnet Search}

The Verifiable Organizations Network has already implemented a basic search functionality\footnote{Verifiable Organizations Network Sovrin mainnet search: https://sovrin-mainnet-browser.vonx.io/browse/domain, accessed on 28.8.2019} for the Sovrin mainnet. 
However, it is lacking essential features that our use case requires. For example, the first transaction is authored by Phil Windley, but if we filter the transactions for "Phil Wimdley" (where the correct way of spelling is "Phil Windley") we get no results as of 28.8.2019. In addition, we cannot use the VONX Sovrin Mainnet Search for retrieving credential definitions by the name of the schema, as this information is not included in the respective ledger transaction. The aforementioned functionalities are already implemented for our full-text search service making it superior to the state of the art for our scenario. 
Furthermore, our use case is also different from that of the Verifiable Organizations Network, as we have integrated it into a verifier application, in order to help the user find schemas and credentials of interest easily. In other words, we imagine ledger search as a frequent part of the SSI workflow.

\subsubsection{Jolocom}

Jolocom \cite{Webpage:Jolocom}, \cite{Whitepaper:Jolocom} is a startup aiming to decentralize identity. The design focuses on usability, security, integrated IoT and human identity, portability, and compliance to standards and best practices. Jolocom is referring to the W3C DID and DDO specification and the BIP 32,39 and 44 standards for the derivation of key pairs.

\subsubsection{uPort}

UPort \cite{Webpage:uPort} is an Ethereum based SSI platform based on DIDs and verifiable credentials. UPort products include a Self-sovereign Wallet, Authentication and Single sign-on (SSO) for decentralized applications, (verifiable) credentials, and soon a mobile SDK as well.

\subsubsection{Veres One}

The Veres One Project \cite{Webpage:VeresOne} is creating the Veres One Network, which does not utilize a potentially scarce and volatile ID token, but charges low and stable fees for avoiding potential regulatory risks of speculative token networks.

\subsubsection{Ocean Protocol}
Ocean Protocol envisions a decentralized data sharing platform unlocking data for artificial intelligence. The startup believes that decentralization would make data on servers accessible for analysis. Ocean protocol's DID method is also part of the planned DID method registry.

\subsubsection{Side Trees and Microsoft ION}
Sidetrees \cite{Webpage:Sidetree} are a relatively scalable and blockchain agnostic Layer 2 protocol for DID operations while maintaining immutability and verifiability. The higher transaction throughput is achieved by batching multiple DID operations into a single blockchain transaction. A Sidetree network consists of Sidetree nodes running the Sidetree protocol, and the consensus algorithm of the underlying blockchain helps serializing the operations committed by different nodes. The actual DID Documents are stored in a Distributed Content Addressable Storage (DCAS or CAS). Running a DCAS node for Sidetree does not require running the Sidetree protocol itself in order to provide redundancy. Microsoft has unvieled ION \cite{GitHub:ION} based on the Sidetree Protocol \cite{Webpage:Sidetree}, which is said to be able to handle tens-of-thousands of transactions \cite{Blog:ION}.

\subsubsection{Alastria}

A National blockchain initiative from Spain backed by multiple large companies from the country is Alastria. The non-profit association is not focusing solely on SSI, but \textit{promotes the digital economy through the development of decentralised ledger technologies/Blockchain}, according to their website\footnote{Alastria Website: https://alastria.io/en/, accessed on 28.8.2019}.

\subsubsection{Civic and Identity.com}

Another prominent member of the SSI movement is Civic \cite{Webpage:Civic}, and the website Identity.com is leveraging \cite{Webpage:Identity} for creating an SSI marketplace via smart contracts facilitating payment. They also offer identity requester, identity validator, and credential wallet toolkit.

\section{Design}
\label{design}

\begin{figure*}[t!]
    \centering
    \includegraphics[width=\textwidth]{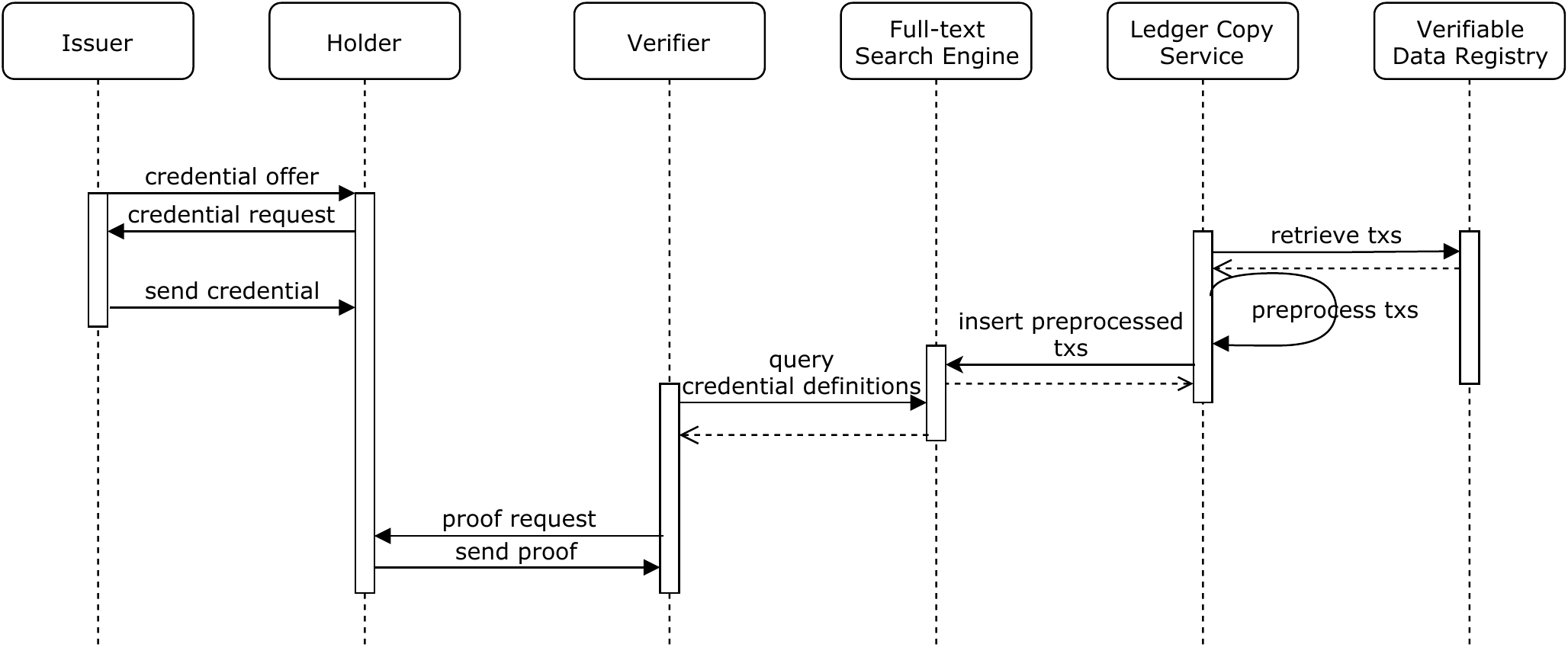}
    \caption{The full-text verifiable credential metadata search process integrated into the SSI workflow. First, the issuer sends a credential offer to the (identity) holder. Then, the holder sends a credential request to the issuer. In the next step of the process the issuer sends the credential to the holder. Then, our proposed full-text search system is used by the verifier for formulating relevant restrictions about trusted credentials. The steps for bootstrapping our full-text search service are as follows: (1) the ledger copy service retrieves transactions from the verifiable data registry, (2) then the ledger copy service preprocesses the records of the verifiable data registry and inserts them into the full-text search engine. Now, the verifier can already query the full-text search engine for finding relevant credential definitions or schemata. Once the proper schemata and credential definitions are selected, the verifier sends a proof request to the holder. Upon receiving the proof request, the holder can decide to respond by sending a cryptographically verifiable proof to the verifier. For the sake of simplicity, we have excluded the steps for establishing pairwise connections between different parties (issuer, holder, and verifier) and verifying the received proof based on the verifiable data registry. The abbreviation \textit{txs} stands for \textit{transactions}.}
    \label{fig:ssi_search_seq}
\end{figure*}

A ledger search tool for an identity use case has to be performant, intuitive, and trusted. Solving the first two criteria can be solved by maintaining a local copy of the ledger in a suitable persistent storage and integrating full-text search software such as Elasticsearch \cite{Webpage:Elastic} or Apache Solr into our software stack (note, that both Apache Solr and Elasticsearch are based on Apache Lucene \cite{Webpage:ApacheLucene}). Fetching transactions from the ledger each time would clearly limit the responsiveness of our system, and it would also make integration to any full-text search software at least cumbersome if not impossible. If Hyperledger Indy along with SSI gets worldwide adoption, there would potentially be millions of schemas and credential definitions on the ledger, proportional to the number of organizations worldwide.

Furthermore, we need a microservice regularly polling the ledger for new transactions and inserting them into a database or another solution with proper storage capabilities. Creating and regularly updating an index could also become useful in the future for achieving low query response times. Using a database for storing the contents of the ledger would be a convenient choice for general purpose usage, and as the records are essentially JSON-LD objects one can think of a popular NoSQL database. On the other hand, full-text search solutions such as Apache Solr or Elasticsearch can also serve as a persistent storage for the local read-only ledger copy.

A high-level overview of integrating the process into the SSI identification workflow is illustrated in Figure \ref{fig:ssi_search_seq}.

A question we should adhere is if we need to keep scalability in mind, when designing a search system for identity on a global scale. The answer is partially yes, as we need to maintain a local copy of the ledger. An option for achieving performance gains by using only relevant transaction types to maintain a separate database for the full-text search. As Hyperledger Indy keeps the vast majority of the data off-chain, we do not need to rely on software with big data capabilities even when Sovrin will be used on a global scale. So in our case sharding is sufficient to address the future storage needs of a full-text search service designed for long term maintainability.

At the moment a simple in-memory full-text search library would also be sufficient, as the Sovrin Main Net currently contains less then 100 schema and credential definitions together, but this is due to the early stage of SSI. However, if SSI gets worldwide adoption, then we would need to be able to handle at least millions of schemas and credential definitions, proportional to the number of legal entities, with potentially numerous attributes, so simply storing and filtering these records would also need additional measures on the backend side.

When designing how to include the full-text search functionality into the backend, more decoupled architectures provide more flexibility, and the complexity of the feature is also in support of not including it into a single monolithic backend along with other SSI and non-SSI related business logic.

Designing the architecture for our full-text search engine also requires understanding the semantics of the ledger transactions, the relationships between them, and our use case. For example, the \texttt{alias} field in \texttt{NYM} transactions is potentially valuable for the full-text search, as it can include a human readable name for the holder of the DID. As the \texttt{alias} field is not present in \texttt{SCHEMA} and \texttt{CLAIM\_DEF} (credential definition) transactions of Hyperledger Indy, we can enrich them with this information for more relevant search results with user input similar to the \texttt{alias} of the user is included in the query, like "Desert Schools Credit Union". Such enhancements require more preprocessing and also program calls for accessing previous records of the ledger to the full-text search engine. \texttt{SCHEMA} transactions also contain a \texttt{name} attribute under the path \texttt{txn.data.data.name} in the JSON-LD document, which is not included in \texttt{CLAIM\_DEF} transactions. But \texttt{CLAIM\_DEF} transactions include a \texttt{ref} attribute under the path \texttt{txn.data.ref} referring to the \texttt{txnMetadata.seqNo} property of the respective \texttt{SCHEMA}, so we can join \texttt{SCHEMA} and \texttt{CLAIM\_DEF} as a preprocessing step in order to enrich \texttt{CLAIM\_DEF} transactions with the name of the \texttt{SCHEMA} they are based on. Thus, we propose using a full-text search engine along with a lightweight component for downloading, preprocessing and enriching the ledger transactions potentially along with an additional database for querying related transactions during the preprocessing stage with the purpose of appending information from related records.

For satisfying the trust criteria we first need to guarantee the integrity of our local copy of the chain, and also make sure that the examined credential is coming from a trusted source. Verifying the integrity of the local copy is easier than satisfying the trust criteria towards the issuer, and this is something that anyone is able to perform. In this paper, we will assume that the \texttt{alias} field in \texttt{NYM} transactions can be relied on for our purposes, and as we are focusing on public permissioned ledgers, namely Hyperledger Indy and Sovrin, we can also assign a certain level of trust towards the information stored on the ledger.

\section{Implementation}
\label{implementation}
An easy and up-to-date way to implement the SSI search service is to package each component as a docker container. The search service used by the frontend can be accessed through a REST API. 

\subsection{Full-text Search Engine}

For a convenient and intuitive search service we need at least full-text search, so the verifiers can conviniently search for schemas and credential definitions on the ledger. We  often want to find credential definitions which contain information we can only roughly specify. For example, we would like to find ID card credential definitions, which might be called simply ID, ID card, personal ID, etc..

\subsubsection{Apache Solr}
We decided to use Apache Solr, as it is an open source and completely free to use software. For commercial purposes one might also consider Elasticsearch or any other software offering a rich full-text search functionality. Apache Solr is also part of the standard Cloudera Hadoop installation targeting big data environments, it supports sharding and replication, so it would be a good solution even if the ledger becomes very large.

\subsubsection{Local ledger Copy}

First, we can examine how others have implemented software that fetches the ledger and persists it in a on-premises storage. The open source software Indyscan is using MongoDB for storing the ledger \cite{Webpage:Indyscan}.
However, maintaining a local database is not crucial for a basic full text search in our case. We can simply add the documents to Solr, for example with REST requests in Javascript.
For a responsive search we need an up-to-date local copy of the examined ledger maintained in a database under our control in our own infrastructure. MongoDB is a convenient choice for storing the Sovrin ledger, as the individual records are valid JSON documents. For connecting to the Sovrin Mainnet we were using our own REST API. By simply providing the pool genesis transaction file we can connect to the Indy ledger, and then we retrieve the ledger transactions for persisting them in a local MongoDB. This can be done by a simple Javascript program, which is periodically polling the ledger for new transactions. In the beginning, the program is requesting transactions as fast as possible, and once the local ledger copy is up to date, it is enough to poll for new transactions with a much lower frequency. During fetching the records we can also enrich and preprocess them in a way that enhances the full-text search process. For example, we can add human readable information about the author of the transaction containing the specific credential definition or schema. 

\subsection{DIMS API}
The DIMS API is a REST API wrapping the functionality of the Indy SDK and simplifying its usage. We have also used it to implement and demonstrate proof of concept SSI issuer and verifier web applications. By using the IDChain/DIMS API we can easily build web applications utilizing DIDs and Verifiable Credentials powered by Hyperledger Indy. Our REST API can also be used to create a local copy of the contents of the Indy ledger. We can also specify ranges of transactions, so we do not need an individual call per transaction, saving most of the network round trip time that would otherwise be needed with one transaction retrieved per query.

As a major focus of Hyperledger Indy is the privacy of the identity holder, which means that private information is primarily stored in the wallet of the user. Our DIMS API uses MongoDB for storing the wallets of the user, which contains credentials, connections and other necessary account information. We also have a cloud agent implementation serving as a persistent endpoint and forwarding messages to the users of our API, which is needed, because end users do not have a fix IP address where they are reachable all the time. Cloud agents with Hyperledger Indy do not need to have the full trust of the identity holder, as they are unable to decrypt the messages they are forwarding to the user.

\subsection{Frontend}

We also need to think about how to implement the search functionality on the UI side. On the frontend side we only need to be able to access the REST API of the chosen full-text search engine. We only need a single HTTP GET request for querying, and in the response the relevant records of the ledger are returned. For the query only a single text field is necessary. On the other hand, depicting the results depends on the use case. In case of the verifiable credential search we should be able to show detailed information about a credential definition upon interest and also select one or multiple credential definitions for creating a proof request.

\section{Evaluation}
\label{evaluation}

Without an intuitive full-text search functionality for the ledger we would either have to rely on hard coded restrictions for credentials like credential definition IDs or a cumbersome way of directly looking up such information at the individual websites of issuers (provided it is mentioned there). Another option would be to manually go through all ledger transactions on a website like indyscan.io. The best alternative at the moment would be to use the VONX Mainnet search \cite{Webpage:VONXBrowser}, but using it would still require understanding the semantics of the Indy ledger, which cannot be required from an end user. Furthermore, one can also not search transactions by relying on the alias of a certain user at the moment. In addition, our ledger search functionality is designed to be integrated into the application a verifier is using, meaning that it is the most convenient option to use according to our current knowledge, as the VONX Mainnet search would involve an additional step. A possibility to look into for simplifying the process without loosing flexibility would be to be able to decide automatically using a complex business logic or machine learning algorithm if a credential credential can be accepted or not.

\subsection{The need for full-text search}
If we simply rely on filtering verifiable credentials, then we need exact matches for specific attributes. With full-text search we can find relevant content even with typos and just partially matching user input. Thus, we can safely assume that a full-text search would significantly improve the user experience and potentially lead to higher recall in retrieving relevant information about credentials found on the ledger.

\subsection{Use cases}

In addition to the use case of the online housing platform mentioned in Section \ref{Introduction}, there are many other promising ways to utilize textual queries for retrieving metadata about credentials stored on the ledger.

We could also think of an online self-service loan on an e-commerce platform, where the identity holder sends himself/herself a proof request for the proof of income certificate credential definition written on the ledger by his employer using user friendly full-text search. In addition, we can think of an interactive usage, when a person (holder) is applying for a loan in a shop, and tells the salesperson (verifier) the name of his/her employer, and the verifier then uses our full-text search service to find information about the proof of employment credential for that specific employer and then send a proof request. 

Furthermore, an employee of the human resources department of an organization would need to be able to ask for a wide variety of credentials like university degrees, ID cards, drivers license, documents about previous employment history, and for the sake of simplicity necessary information about credentials could be retrieved by using full-text search on the ledger. A self-service version of the previous scenario is when the identity holder logs in at the webpage of the organization and searches for credentials issued by a specific issuer from a potentially very large pool of trusted candidates, like an attestation of sickness from any medical doctor, and then sends a proof request to herself. Another example is, when an IT person configuring and maintaining an Indy edge agent verifier software for a specific client could search for trusted credential definitions and issuers.

\subsection{Potential Users of the Full-text Search}

The full-text search can be employed for a variety of worfklows and use cases. This paper mainly focuses on the employing full text search on the verifier or the issuer side.

\subsection{Need for Updating the List of accepted Credential Types}
The list of issuers is continuously growing, thus verifiers would often need to be informed about new issuers. For example, in case of a proof of employment credential, each employer can be an issuer. Furthermore, newer versions of the schema would also need to be found. A possible way for maintaining an up to date list of relevant credential definitions, schemas and issuers for a proof request is to regularly search for new relevant content on our local copy of the ledger, which could be performed by the operator/maintainer of an application with verification functionality.

\subsection{The Potential Size of the Pool of Credential Types}

Note, that the list of potential employers world wide is much too long to include in an ordinary proof request. According to datapo \cite{Webpage:NumCompanies}, there are millions of companies and entrepreneurs registered worldwide, meaning that a comprehensive proof of employment list is highly impractical or not even feasible to send in practice in a proof request, so if Hyperledger Indy gets worldwide adoption, we would either need to be able to find a likely relevant subset of credential definitions by using additional information, or we could rely on proof offers from the identity holder and that every entity writing credential definitions on the permissioned ledger can be trusted. Note, that at the moment, we did not find an implementation for proof offers with the Hyperledger Indy SDK, and these are included in a Hyperledger Aries RFC \cite{GitHub:HyperledgerAriesRFC}, which is in accepted status as of September 2, 2019. As Hyperledger Indy is a permissioned ledger, where the right to be able to write on the ledger (the role of trust anchor) can be granted by Sovrin stewards, the question is if verifiers trust them. If complete blind, trust is the case, then verifiers could also just send proof requests with no restrictions about credential definitions. However, we are still skeptical if each verifier would accept accept any credential. Thus, two main options are: 1) the identity holder tells the verifier about his employer's proof of employment certificate, or 2) the verifier maintains a list of potential credential definition IDs that include a large portion of the relevant records.

Still, even in this scenario, issuers would need to be able to find relevant schemas and credential definitions on the ledger, in order to be able to issue credentials with a set of attributes that are widely accepted by verifiers. In order to fulfill this requirement, we would either need use case specific schemas that issuers implicitly agree on, or another way to be able to have a wide agreement on attribute names among issuers. For example, with a proof of employment credential we cloud agree on the attributes: name, company and title. Now the question is how will issuers be able to specify matching attribute names, and our answer is that the easiest way is to look at the ledger, as all credential definitions and schemas and stored on the blockchain in case of Hyperledger Indy. For being able to find relevant schemas and credential definitions easily, issuers can rely on our full-text search service.

Alternatives to our search service could be either trying to find information online or be informed on relevant attribute names by use case dependent third parties, such as a government in case of a proof of employment credential. Finding information online would be an extra overhead, and it would be much better for the user experience on the issuer side, if the attribute selection could be solved by a full-text ledger search functionality by either finding a proper schema for the credential definition to be created, or by retrieving a set of schema and credential definitions that can serve as a basis for specifying the right attribute names.

\subsection{Interactive Initiation of the Identification Process on the Holder Side}
Another valid question is if the identity holder just tell the Verifier the credential definition ID.
The credential definition ID is not human readable, so even if it is sent in an automated way, the verifier will have to look the respective credential definition on the ledger and decide if that is appropriate for the specific scenario. So even with a proof offer, the verifier will have to look at the ledger and decide if the suggested credential is appropriate. A proof offer could be an alternative for the verifiable credential search in a scenario where the identity holder and verifier persons are negotiating the proof in an interactive manner, as long as the identity holder only has a number of credentials that he/she can manually go through. If the identity holder has a hundreds of credentials, then he/she would also need to rely on a search functionality, but this time the place to search would be the own wallet.

\subsection{Obtaining a Matching Credential}

Furthermore, it is also possible that the identity holder does not have a matching credentials, which means he/she will need to obtain one from a list of possible credential types. For being able to select the credential the holder will try to acquire simple filtering is also sufficient as long as the number of matching credential definitions (according to Hyperledger Indy terminology) is not too long, because the holder can manually go through the list of accepted credentials. If the list is too long, then we will again need an intuitive full-text search functionality for the convenience of the user.

\subsection{Trust}

Trust is generally a central question when working with blockchains, and our case is no exception. With SSI the authenticity of verifiable credentials can be cryptographically verified, but then trust in the issuer of the credentials is needed. This aligns well with the notion described in the paper authored by Heiss \cite{heissoracles} that contents on the chain need to be trusted. In the case of Hyperledger Indy, the Sovrin foundation has to make sure, that trust anchors are indeed widely trusted entities by society when it comes to credentials they issue. As Sovrin stewards are already a large consortium of reputable organizations, we can safely assume that they have the necessary capacity to evaluate which organizations are qualified to become trust anchors. If we were able to trust issuers in general, this could save us tremendous efforts by reducing the need to formulate proper restrictions on which credentials can automatically be trusted, and the business processes with Hyperledger Indy could be automatized to a higher extent.

\subsection{Security}

First, we would like to emphasize again that no private data is stored on the chain with Hyperledger Indy even in encrypted form. In this paper, we only search for information that is meant to be public, for example schemas, credential definitions and public DIDs of issuers.

For the security of the user, we only need to make sure, that the local ledger copy is not tampered with. Once we have access to the genesis transaction, we can easily perform this upon receiving the transactions in the order they were created.

As the full-text ledger search can also be employed for verifying information about credentials displayed on other media, we can state that it can be beneficial for improving the security of applications utilizing SSI.

\subsection{Performance}

\begin{figure}[t!]
\centerline{\includegraphics[width=\linewidth]{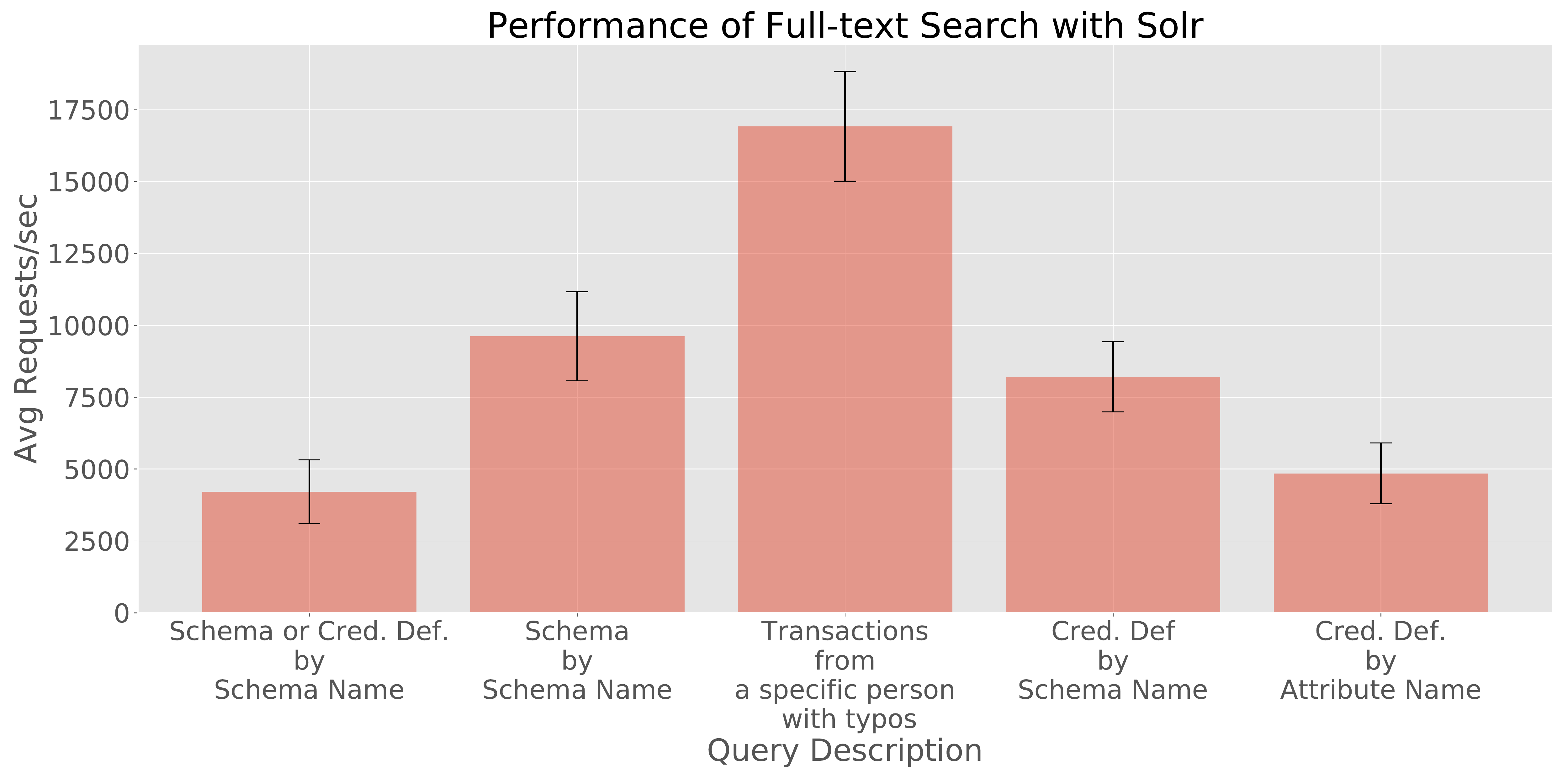}}
\caption{The performance of our full-text search service powered by Solr (single node deployment on commodity hardware). The throughput of the search service has been measured over varying query types.}
\label{fig:solr_performance}
\end{figure}

We have conducted performance benchmarks of our full-text search service deployed on a commodity machine. We have tested the throughput using the \texttt{wrk} package with using 400 open connections and a single thread. The performance of five query types were evaluated, namely querying  1) schemas or credential definitions by schema name, 2) schema by schema name, 3) transactions from a specific person with a typo in the name, 4) credential definition by schema name, and 5) credential definition by attribute name. Our local Solr instance contained the first 47312 domain transactions of the Sovrin main net. Our current benchmarks show that a single node Solr setup can serve up to 16000 requests per second depending on the type of query. Thus, we can conclude that our full-text search service is more performant than necessary. The performance metrics could significantly change in the future, as SSI is currently in an early stage, and the Sovrin ledger will contain more transactions by potentially orders of magnitude if they achieve worldwide adoption. The results of our measurements are depicted in Figure \ref{fig:solr_performance}. 

Note that Apache Solr indexes can be replicated over multiple nodes, which lets us safely assume linear performance scaling. Solr also offers index sharding, which could be useful if the Sovrin ledger becomes too large. The scaling of sharded indexes can also be achieved by manually replicating them across several Solr clusters.

\section{Conclusion}
\label{conclusion}

We have envisioned how a verifiable credentials based ecosystem could be functioning on a global scale through the means of Self-sovereign Identity, where in our scenario a full-text search feature helps the users of the system retrieve necessary information for trusted and automatic verification. We have implemented and evaluated a prototype full-text credential metadata search service, which serves as a proof of our concept. Our text based search engine could help verifiers find credential types relevant for their use-cases created by trusted issuers among millions of potential candidates, which would be a requirement for a worldwide used verifiable credential system.

\subsection{Future Work}
Interoperability is a major requirement for SSI, and the standardization by the World Wide Web Consortium has the promise of helping it. Potential next steps could be extending the full text search functionality to support uPort, Jolocom, and other SSI software as well. Furthermore, the Hyperledger Aries project \cite{Webpage:Aries} also aims to achieve interoperability with verifiable credentials.

\subsubsection{Integrity verification of the chain}
For prooving that we are searching on an untampered version of the chain we need to check its integrity. In order to verify the blockchain we need to recompute the hashes of the transactions. When using Hyperledger Indy, we would have to follow the computationally efficient Merkle tree specific verification process. As we are periodically polling the ledger, we can integrate the verification process into the regular update mechanism looking for new transactions.

\subsubsection{Schema Overlays}
The idea of using overlays \cite{Webpage:SchemaOverlays} has emerged in the Hyperledger Indy community. Overlays would provide additional contextual and conditional information about schemas for SSI capable applications and services, and they would be a rich source to incorporate in our full-text search, once they are available and used.

\subsection{Final Word}

In conclusion, full-text search is a useful and promising tool for building a verifiable credential ecosystem that can easily and flexibly play along with other approaches in creating an efficient verification process between holders and verifiers.

\section*{Acknowledgment}

We would like to thank {\"O}mer Ilhan, Santhos Baala Ramalingam Santhanakrishnan, and Marcel Ebermann for their work within the DIMS project\footnote{DIMS Project: https://www.snet.tu-berlin.de/menue/projects/dims/, accessed on 28.8.2019} on working with Self-sovereign Identity, Verifiable Credentials, and Decentralized Identifiers.

\bibliographystyle{./IEEEtran}
\bibliography{./IEEEabrv,./IEEEIndy}

\begin{thebibliography}{10}
\providecommand{\url}[1]{#1}
\csname url@samestyle\endcsname
\providecommand{\newblock}{\relax}
\providecommand{\bibinfo}[2]{#2}
\providecommand{\BIBentrySTDinterwordspacing}{\spaceskip=0pt\relax}
\providecommand{\BIBentryALTinterwordstretchfactor}{4}
\providecommand{\BIBentryALTinterwordspacing}{\spaceskip=\fontdimen2\font plus
\BIBentryALTinterwordstretchfactor\fontdimen3\font minus
  \fontdimen4\font\relax}
\providecommand{\BIBforeignlanguage}[2]{{%
\expandafter\ifx\csname l@#1\endcsname\relax
\typeout{** WARNING: IEEEtran.bst: No hyphenation pattern has been}%
\typeout{** loaded for the language `#1'. Using the pattern for}%
\typeout{** the default language instead.}%
\else
\language=\csname l@#1\endcsname
\fi
#2}}
\providecommand{\BIBdecl}{\relax}
\BIBdecl

\bibitem{Whitepaper:Sovrin}
\BIBentryALTinterwordspacing
Sovrin whitepaper. [Online]. Available:
  \url{https://sovrin.org/wp-content/uploads/Sovrin-Protocol-and-Token-White-Paper.pdf}
\BIBentrySTDinterwordspacing

\bibitem{allen2016path}
\BIBentryALTinterwordspacing
C.~Allen, ``{The Path to Self-Sovereign Identity},'' \emph{{Life with
  Alacrity}}, 2016. [Online]. Available:
  \url{http://www.lifewithalacrity.com/2016/04/the-path-to-self-soverereign-identity.html}
\BIBentrySTDinterwordspacing

\bibitem{tobin2016inevitable}
A.~Tobin and D.~Reed, ``{The Inevitable Rise of Self-Sovereign Identity},''
  \emph{The Sovrin Foundation}, vol.~29, 2016.

\bibitem{Webpage:VC}
\BIBentryALTinterwordspacing
Verifiable credentials. [Online]. Available:
  \url{https://www.w3.org/TR/verifiable-claims-data-model/}
\BIBentrySTDinterwordspacing

\bibitem{Webpage:ApacheSolr}
\BIBentryALTinterwordspacing
Apache solr. [Online]. Available: \url{https://lucene.apache.org/solr/}
\BIBentrySTDinterwordspacing

\bibitem{nakamoto2008bitcoin}
S.~Nakamoto \emph{et~al.}, ``Bitcoin: A peer-to-peer electronic cash system,''
  2008.

\bibitem{wood2014ethereum}
G.~Wood \emph{et~al.}, ``Ethereum: A secure decentralised generalised
  transaction ledger,'' \emph{Ethereum project yellow paper}, vol. 151, pp.
  1--32, 2014.

\bibitem{dunphy2018first}
P.~Dunphy and F.~A. Petitcolas, ``A first look at identity management schemes
  on the blockchain,'' \emph{IEEE Security \& Privacy}, vol.~16, no.~4, pp.
  20--29, 2018.

\bibitem{wilcox2003names}
Z.~Wilcox-O'Hearn, ``{Names: Decentralized, Secure, Human-Meaningful: Choose
  Two},'' \emph{online] https://web. archive.
  org/web/20011020191610/http://zooko.com/distnames.html [retrieved
  2018-04-21]}, 2003.

\bibitem{fromknecht2014decentralized}
C.~Fromknecht, D.~Velicanu, and S.~Yakoubov, ``{A Decentralized Public Key
  Infrastructure with Identity Retention.}'' \emph{IACR Cryptology ePrint
  Archive}, vol. 2014, p. 803, 2014.

\bibitem{allen2015decentralized}
C.~Allen, A.~Brock, V.~Buterin, J.~Callas, D.~Dorje, C.~Lundkvist,
  P.~Kravchenko, J.~Nelson, D.~Reed, M.~Sabadello \emph{et~al.},
  ``{Decentralized Public Key Infrastructure. A White Paper from Rebooting the
  Web of Trust},'' 2015.

\bibitem{garman2014decentralized}
C.~Garman, M.~Green, and I.~Miers, ``{Decentralized Anonymous Credentials},''
  in \emph{NDSS}.\hskip 1em plus 0.5em minus 0.4em\relax Citeseer, 2014.

\bibitem{Webpage:DIF}
\BIBentryALTinterwordspacing
Decentralized identity foundation. [Online]. Available:
  \url{https://identity.foundation}
\BIBentrySTDinterwordspacing

\bibitem{Webpage:VCUseCases}
\BIBentryALTinterwordspacing
Verifiable claims use cases. [Online]. Available:
  \url{https://www.w3.org/TR/verifiable-claims-use-cases/}
\BIBentrySTDinterwordspacing

\bibitem{Webpage:BUDRU}
\BIBentryALTinterwordspacing
(2019) Demo b\"{u}rgeramt. [Online]. Available:
  \url{https://bcp-ssi.tir.budru.de}
\BIBentrySTDinterwordspacing

\bibitem{muhle2018survey}
A.~M{\"u}hle, A.~Gr{\"u}ner, T.~Gayvoronskaya, and C.~Meinel, ``{A Survey on
  Essential Components of a Self-Sovereign Identity},'' \emph{Computer Science
  Review}, vol.~30, pp. 80--86, 2018.

\bibitem{Webpage:DID}
\BIBentryALTinterwordspacing
Decentralized identifiers. [Online]. Available:
  \url{https://w3c-ccg.github.io/did-spec/}
\BIBentrySTDinterwordspacing

\bibitem{Webpage:DIDMethods}
\BIBentryALTinterwordspacing
W3c did method registry. [Online]. Available:
  \url{https://w3c-ccg.github.io/did-method-registry/}
\BIBentrySTDinterwordspacing

\bibitem{Webpage:Jolocom}
\BIBentryALTinterwordspacing
Jolocom. [Online]. Available: \url{https://jolocom.io/}
\BIBentrySTDinterwordspacing

\bibitem{Webpage:OceanProtocol}
\BIBentryALTinterwordspacing
Ocean protocol. [Online]. Available: \url{https://oceanprotocol.com/}
\BIBentrySTDinterwordspacing

\bibitem{Webpage:Uniresolver}
\BIBentryALTinterwordspacing
Uniresolver. [Online]. Available: \url{https://uniresolver.io/}
\BIBentrySTDinterwordspacing

\bibitem{GitHub:Uniresolver}
\BIBentryALTinterwordspacing
Uniresolver repository. [Online]. Available:
  \url{https://github.com/decentralized-identity/universal-resolver/}
\BIBentrySTDinterwordspacing

\bibitem{Webpage:Uniregistrar}
\BIBentryALTinterwordspacing
Uniregistrar. [Online]. Available: \url{https://uniregistrar.io/}
\BIBentrySTDinterwordspacing

\bibitem{Webpage:DIDEverything}
\BIBentryALTinterwordspacing
S.~Conway, A.~Hughes, M.~Ma, J.~Poole, M.~Riedel, S.~M. Smith, and
  C.~St{\"o}cker. (2019) {A DID for Everything}. [Online]. Available:
  \url{https://nbviewer.jupyter.org/github/WebOfTrustInfo/rwot7-toronto/blob/master/final-documents/A_DID_for_everything.pdf}
\BIBentrySTDinterwordspacing

\bibitem{gondor2016distributed}
S.~G{\"o}nd{\"o}r, F.~Beierle, S.~Sharhan, and A.~K{\"u}pper, ``{Distributed
  and Domain-Independent Identity Management for User Profiles in the SONIC
  Online Social Network Federation},'' in \emph{International Conference on
  Computational Social Networks}.\hskip 1em plus 0.5em minus 0.4em\relax
  Springer, 2016, pp. 226--238.

\bibitem{gondor2018seamless}
S.~G\"{o}nd\"{o}r, ``{Seamless Interoperability and Data Portability in the
  Social Web for Facilitating an Open and Heterogeneous Online Social Network
  Federation},'' Ph.D. dissertation, Technische Universit\"{a}t Berlin, 2018.

\bibitem{Webpage:Namecoin}
\BIBentryALTinterwordspacing
{Namecoin}. [Online]. Available: \url{https://www.namecoin.org/}
\BIBentrySTDinterwordspacing

\bibitem{ali2016blockstack}
M.~Ali, J.~Nelson, R.~Shea, and M.~J. Freedman, ``{Blockstack: A Global Naming
  and Storage System Secured by Blockchains},'' in \emph{2016 USENIX Annual
  Technical Conference (USENIX 16)}, 2016, pp. 181--194.

\bibitem{Webpage:Sovrin}
\BIBentryALTinterwordspacing
Sovrin. [Online]. Available: \url{https://sovrin.org/}
\BIBentrySTDinterwordspacing

\bibitem{khovratovich2017sovrin}
D.~Khovratovich and J.~Law, ``Sovrin: digital identities in the blockchain
  era,'' \emph{Github Commit by jasonalaw October}, vol.~17, 2017.

\bibitem{GitHub:Indy}
\BIBentryALTinterwordspacing
Hyperledger indy github. [Online]. Available:
  \url{https://github.com/hyperledger/indy-sdk}
\BIBentrySTDinterwordspacing

\bibitem{Webpage:Aries}
\BIBentryALTinterwordspacing
(2019) Hyperledger aries project. [Online]. Available:
  \url{https://www.hyperledger.org/projects/aries}
\BIBentrySTDinterwordspacing

\bibitem{Webpage:Ursa}
\BIBentryALTinterwordspacing
Hyperledger indy. [Online]. Available:
  \url{https://www.hyperledger.org/projects/ursa}
\BIBentrySTDinterwordspacing

\bibitem{lamport1982byzantine}
L.~Lamport, R.~Shostak, and M.~Pease, ``The byzantine generals problem,''
  \emph{ACM Transactions on Programming Languages and Systems (TOPLAS)},
  vol.~4, no.~3, pp. 382--401, 1982.

\bibitem{Webpage:Plenum}
\BIBentryALTinterwordspacing
(2015) Plenum: Byzantine fault tolerant protocol. [Online]. Available:
  \url{https://github.com/evernym/plenum}
\BIBentrySTDinterwordspacing

\bibitem{castro1999practical}
M.~Castro, B.~Liskov \emph{et~al.}, ``Practical byzantine fault tolerance,'' in
  \emph{OSDI}, vol.~99, 1999, pp. 173--186.

\bibitem{buchman2016tendermint}
E.~Buchman, ``Tendermint: Byzantine fault tolerance in the age of
  blockchains,'' Ph.D. dissertation, 2016.

\bibitem{Webpage:VONX}
\BIBentryALTinterwordspacing
Verifiable orginazations network. [Online]. Available: \url{https://vonx.io/}
\BIBentrySTDinterwordspacing

\bibitem{Webpage:Orgbook}
\BIBentryALTinterwordspacing
(2019) {Orgbook BC}. [Online]. Available:
  \url{https://orgbook.gov.bc.ca/en/home}
\BIBentrySTDinterwordspacing

\bibitem{Whitepaper:Jolocom}
\BIBentryALTinterwordspacing
Jolocom. (2018) {Jolocom Whitepaper}. [Online]. Available:
  \url{https://jolocom.io/wp-content/uploads/2018/07/Jolocom-Technical-WP-Self-Sovereign-and-Decentralised-Identity-By-Design
  -2018-03-09.pdf}
\BIBentrySTDinterwordspacing

\bibitem{Webpage:uPort}
\BIBentryALTinterwordspacing
uport. [Online]. Available: \url{https://www.uport.me/}
\BIBentrySTDinterwordspacing

\bibitem{Webpage:VeresOne}
\BIBentryALTinterwordspacing
Veres one. [Online]. Available: \url{https://veres.one/}
\BIBentrySTDinterwordspacing

\bibitem{Webpage:Sidetree}
\BIBentryALTinterwordspacing
Sidetree protocol. [Online]. Available:
  \url{https://github.com/decentralized-identity/sidetree/blob/master/docs/protocol.md}
\BIBentrySTDinterwordspacing

\bibitem{GitHub:ION}
\BIBentryALTinterwordspacing
Microsoft ion. [Online]. Available:
  \url{https://github.com/decentralized-identity/ion}
\BIBentrySTDinterwordspacing

\bibitem{Blog:ION}
\BIBentryALTinterwordspacing
(2019) Microsoft blog, ion announcement. [Online]. Available:
  \url{https://techcommunity.microsoft.com/t5/Azure-Active-Directory-Identity/Toward-scalable-decentralized-identifier-systems/ba-p/560168}
\BIBentrySTDinterwordspacing

\bibitem{Webpage:Civic}
\BIBentryALTinterwordspacing
Civic. [Online]. Available: \url{https://www.civic.com}
\BIBentrySTDinterwordspacing

\bibitem{Webpage:Identity}
\BIBentryALTinterwordspacing
Identity.com. [Online]. Available: \url{https://www.identity.com}
\BIBentrySTDinterwordspacing

\bibitem{Webpage:Elastic}
\BIBentryALTinterwordspacing
Elastic. [Online]. Available: \url{https://www.elastic.co}
\BIBentrySTDinterwordspacing

\bibitem{Webpage:ApacheLucene}
\BIBentryALTinterwordspacing
Apache lucene. [Online]. Available: \url{https://lucene.apache.org/index.html}
\BIBentrySTDinterwordspacing

\bibitem{Webpage:Indyscan}
\BIBentryALTinterwordspacing
Indyscan. [Online]. Available: \url{https://indyscan.io/}
\BIBentrySTDinterwordspacing

\bibitem{Webpage:VONXBrowser}
\BIBentryALTinterwordspacing
Vonx sovrin mainnet search. [Online]. Available:
  \url{https://sovrin-mainnet-browser.vonx.io/browse/domain}
\BIBentrySTDinterwordspacing

\bibitem{Webpage:NumCompanies}
\BIBentryALTinterwordspacing
(2019) How many companies are there in the world? [Online]. Available:
  \url{https://datapo.com/news/how-many-companies-are-there-in-the-world/}
\BIBentrySTDinterwordspacing

\bibitem{GitHub:HyperledgerAriesRFC}
\BIBentryALTinterwordspacing
(2019) {Hyperledger Aries RFC 0037: Present Proof Protocol 1.0}. [Online].
  Available:
  \url{https://github.com/hyperledger/aries-rfcs/tree/master/features/0037-present-proof#presentation-preview}
\BIBentrySTDinterwordspacing

\bibitem{heissoracles}
J.~Heiss, J.~Eberhardt, and S.~Tai, ``From oracles to trustworthy data
  on-chaining systems,'' in \emph{Proc. {IEEE} International Conference on
  Blockchain ({ICBC}2019)}, 2019.

\bibitem{Webpage:SchemaOverlays}
\BIBentryALTinterwordspacing
(2019) {Schema Overlays}. [Online]. Available:
  \url{https://ssimeetup.org/overlays-1o1-establishing-schema-definitions-self-sovereign-identity-ssi-ecosystem-paul-knowles-webinar-17/}
\BIBentrySTDinterwordspacing

\end{thebibliography}

\vspace{12pt}
\color{red}

\includepdf[pages=-]{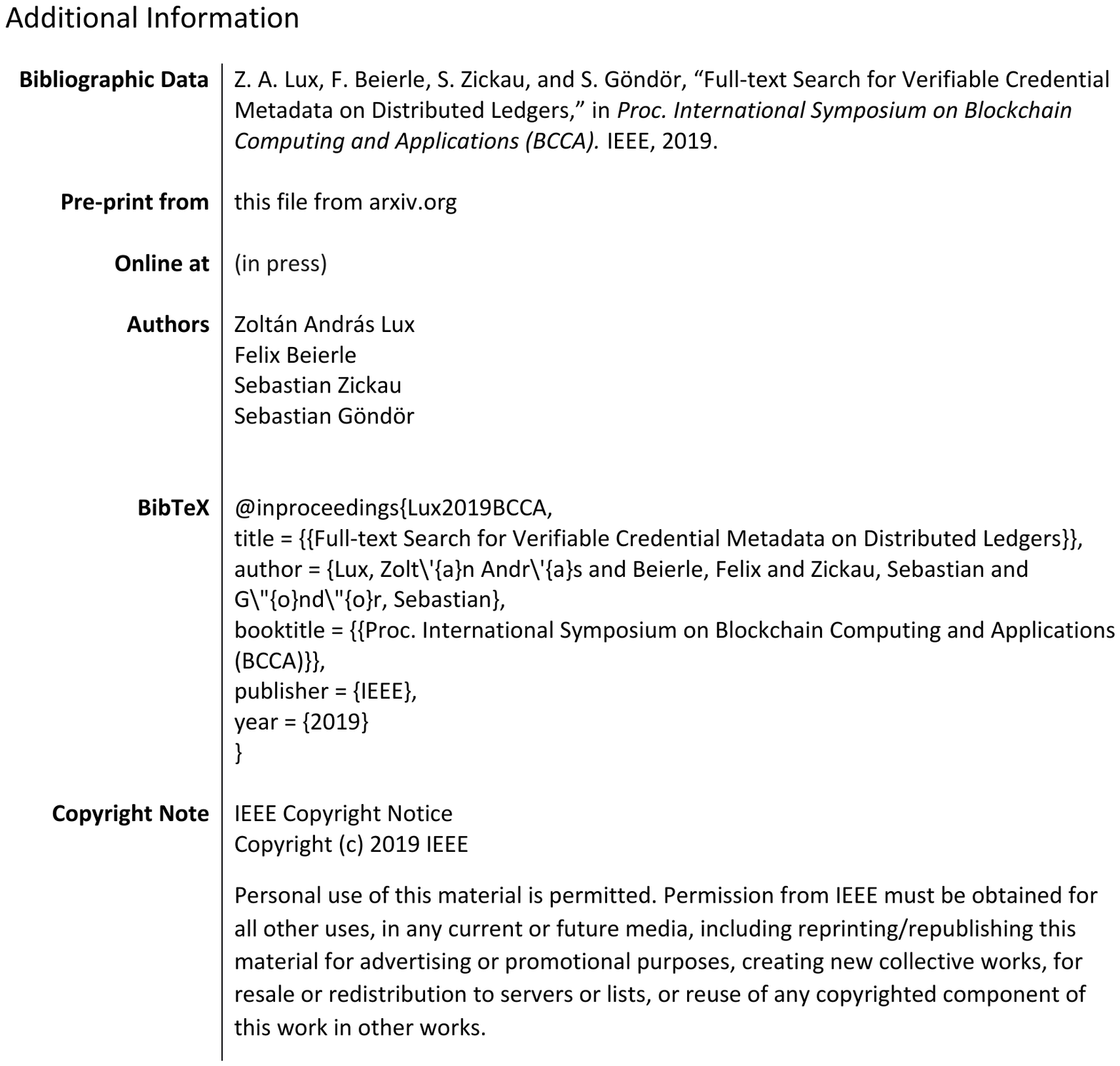}

\end{document}